\documentclass[conference]{IEEEtran}
\usepackage[noadjust]{cite}  
\usepackage{color}
\usepackage{graphicx}
\usepackage{wrapfig}
\usepackage{lipsum}
\usepackage{dblfloatfix}
\usepackage{fixltx2e}
\usepackage{nameref}
\usepackage{amsmath}
\usepackage{graphicx}
\usepackage{varioref}
\usepackage{booktabs}  
\usepackage{siunitx}
\usepackage{numprint}
\usepackage{filecontents}
\usepackage[bookmarks=false]{hyperref}
\ifCLASSINFOpdf
\else
\fi
\IEEEoverridecommandlockouts 
\begin{document}
\title{BigEAR: Inferring the Ambient and Emotional Correlates from Smartphone-based Acoustic Big Data\\ }
\author{\IEEEauthorblockN{Harishchandra Dubey\thanks{\textcolor{blue}{This material is presented to ensure timely dissemination of scholarly and technical work. Copyright and all rights therein are retained by the authors or by the respective copyright holders. The original citation of this paper is:
H. Dubey, M. R. Mehl, K. Mankodiya, BigEAR: Inferring the Ambient and Emotional Correlates from Smartphone-based Acoustic Big Data, IEEE International Workshop on Big Data Analytics for Smart and Connected Health 2016, June 27, 2016, Washington DC, USA.}}\IEEEauthorrefmark{1}, Matthias R. Mehl\IEEEauthorrefmark{2}, Kunal Mankodiya\IEEEauthorrefmark{1}}
\IEEEauthorblockA{\IEEEauthorrefmark{1}Department of Electrical, Computer and Biomedical Engineering, University of Rhode Island, Kingston, RI, USA\\ \IEEEauthorrefmark{2}Department of Psychology, University of Arizona, Tucson, AZ, USA}\\
dubey@ele.uri.edu, mehl@email.arizona.edu, kunalm@uri.edu}

\maketitle
\begin{abstract}
This paper presents a novel BigEAR big data framework that employs psychological audio processing chain (PAPC) to process smartphone-based acoustic big data collected when the user performs social conversations in naturalistic scenarios. The overarching goal of BigEAR is to identify moods of the wearer from various activities such as laughing, singing, crying, arguing, and sighing. These annotations are based on ground truth relevant for psychologists who intend to monitor/infer the social context of individuals coping with breast cancer. We pursued a case study on couples coping with breast cancer to know how the conversations affect emotional and social well being. In the state-of-the-art methods, psychologists and their team have to hear the audio recordings for making these inferences by subjective evaluations that not only are time-consuming and costly, but also demand manual data coding for thousands of audio files. The BigEAR framework automates the audio analysis. We computed the accuracy of BigEAR with respect to the ground truth obtained from a human rater. Our approach yielded overall average accuracy of 88.76\% on real-world data from couples coping with breast cancer. 
\end{abstract}
\IEEEpeerreviewmaketitle
\vspace{-1mm}
\section{Introduction}
Big data have revolutionize all areas of everyday life such as healthcare, psychology, psychiatry~\cite{mayer2013big,pennebaker2003psychological,dubey2015fog}. Automated speech recognition has become a part of our life that involves constant interactions with smartphones. However, speech emotion recognition is a challenging problem, especially when acoustic big data is filling up the cloud storage at a rapid rate. For example, our previous work, known as iEAR~\cite{mehl2007empirical}, recorded a monumental acoustic data from iPhones that involve social and daily life conversations of couples coping with breast cancer. The physiological state of the individuals coping with breast cancer affects their emotions and intimacy that is reflected in everyday conversations. For instance, the mood of such individuals at various times in a day can be inferred from their daily audio log (DAL). Collection of audio data was automated that facilitated unbiased capture of ambient sounds in naturalistic settings. The psychologists had an access to the audio data and also the estimated mood labels that can help in making clinical decisions about the psychological health of the participants. Recently, our team developed a novel framework named as BigEAR that employs a psychological audio processing chain (PAPC) to predict the mood of the speaker from DALs. The BigEAR application installed on the iPhone/iPad captures the audio data from ambient environment at random instance throughout the day. The snippets are of small sizes so that we do not record significant duration of a personal context. The acquired audio data contain speech, music as well as silence. The audio data contain rich information about the emotional correlates in the social conversation such as happiness, sadness or anger.
The presented work makes the following contributions to the area of big data science and analytics: 
\begin{itemize}
\item Development and validation of psychological audio processing chain (PAPC) to generate big-data-driven knowledge about the emotion correlates;
\item Proposed novel perceptual features such as frequency modulation, frequency range, and sharpness for identifying the emotions from speech; 
\item Use of advanced tools for speaker diarization (segmentation) on psychological audio big data;
\item Automating the process of emotion identification from audio data in context of couples coping with breast cancer.
\end{itemize}
\vspace{-2mm}
\section{Background \& Related Works}
Identifying the surrounding environment and emotional
correlates of the naturalistic conversation between individuals is
an important part of the psychologist’s treatment regime for
couples coping with breast cancer~\cite{mets2012naturalistically},~\cite{robbins2014cancer},~\cite{martino2015linguistic}. However, all our previous studies and similar studies by others involved manual transcription and analysis of audio by subjective hearing to decide ambient and emotional correlates for couples  coping with breast cancer~\cite{cheng2014perceived}. 
\begin{figure*}[!t]
\centering
\includegraphics[width=480bp]{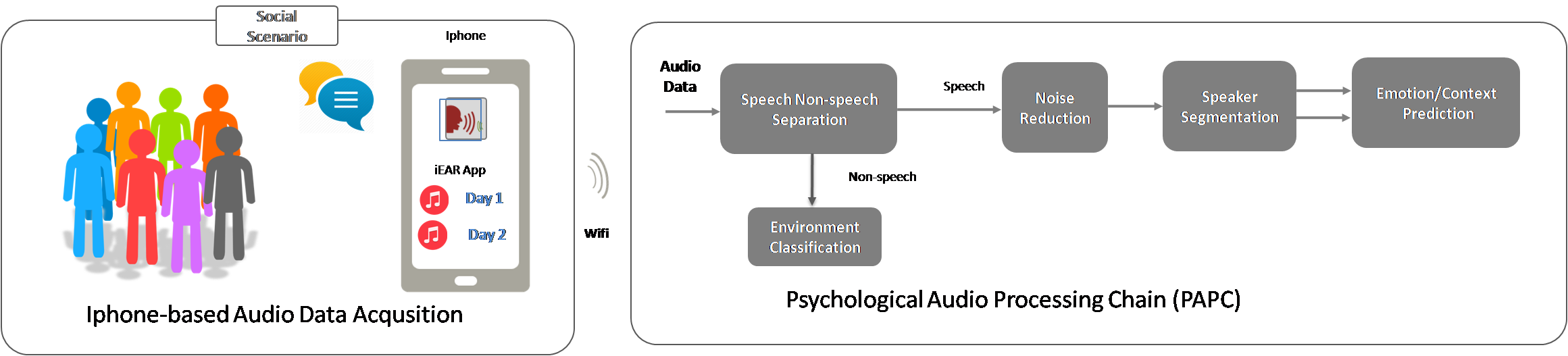}
\caption{A conceptual view of the BigEAR framework. The iEAR application installed on the iPhone captures the audio data from users. The acquired audio data contain speech and non-speech components. The speech data contain rich information about the emotional correlates such as happiness, sadness or neutral emotion while non-speech contain rich information about the surrounding environment such as outdoor, indoor, and TV/music. The audio data were processed by a psychological audio processing chain (PAPC) developed by our team to predict the emotion correlates and identify the surrounding environment. Collection of audio data was automated that facilitated unbiased capture of ambient sounds in naturalistic settings.}
\label{fig_papc}
\end{figure*}
\vspace{0mm}
%
The increasing use of smartphones, smartwatches and wrist-bands provides a convenient interface for collection of audio data without need of any specialized hardware~\cite{dubey2015echowear,dubey2015multi,dubey2015motor,dubey2015fog,monteiro2016fit}. We designed iEAR that is an iPhone app used for unobtrusive collection of ambient sound from user’s surroundings. There are some similar systems in literature such as authors in~\cite{blum2006insense} developed a smart system for collecting the real-life experiences of peoples and store it into an adaptive multimedia bank with variable size. The system used a camera, a microphone and an accelerometer for combined collection of audio and video data when users perform interesting activities. There was a sub-block that detects the presence of interesting activity~\cite{blum2006insense}. However, it required users to use a specialized
hardware and wear it for data collection. In contrast, iEAR does not
require any specialized hardware.
We have used iEAR for audio collection in various research problems in social psychology. For example, we studied self-awareness in individuals using iEAR recorder followed by manual transcription~\cite{mehl2012naturalistic},~\cite{robbins2014cancer}. The individual beliefs regarding their behavior were compared with the inferences brought out by the team of psychologists and clinicians from the collected audio data. The conformance between two metrics quantifies the self-knowledge individual possesses about themselves. We concluded that self-knowledge was positively correlated with the individual's perception of relationship quality. Since the quality of our relationships greatly affects our social life, individual's self-knowledge can add value to their life~\cite{tenney2013examined}. We had devised a method for assessment of social well being and health of individual using iEAR and manual transcription. Such approaches are useful for health researchers and psychologists. iEAR allows social health scientists to hear the audio snippets from individual's personal and social life and hence cross-validate research findings against self-reports by individuals~\cite{mehl2007eavesdropping}. 
In this paper we focus on automated audio analysis-based inference on audio collected by iEAR during social interactions of couples coping with breast cancer. From the ambient audio data, we can know the mood of the couples. The moods are categorical labels that quantify the context of conversation, intensity and type of emotions and sentiment linked to the conversation. The features are usually fed to a classification systems for predicting the class of given pattern classification problem~\cite{dubey2012blind,dubey2012novel}. At the end of this article, we will see the benefits of BigEAR framework for automated psychological audio processing thus reducing logistics and manpower needs. 
\begin{figure}[!t]
\centering
\includegraphics[width=240bp]{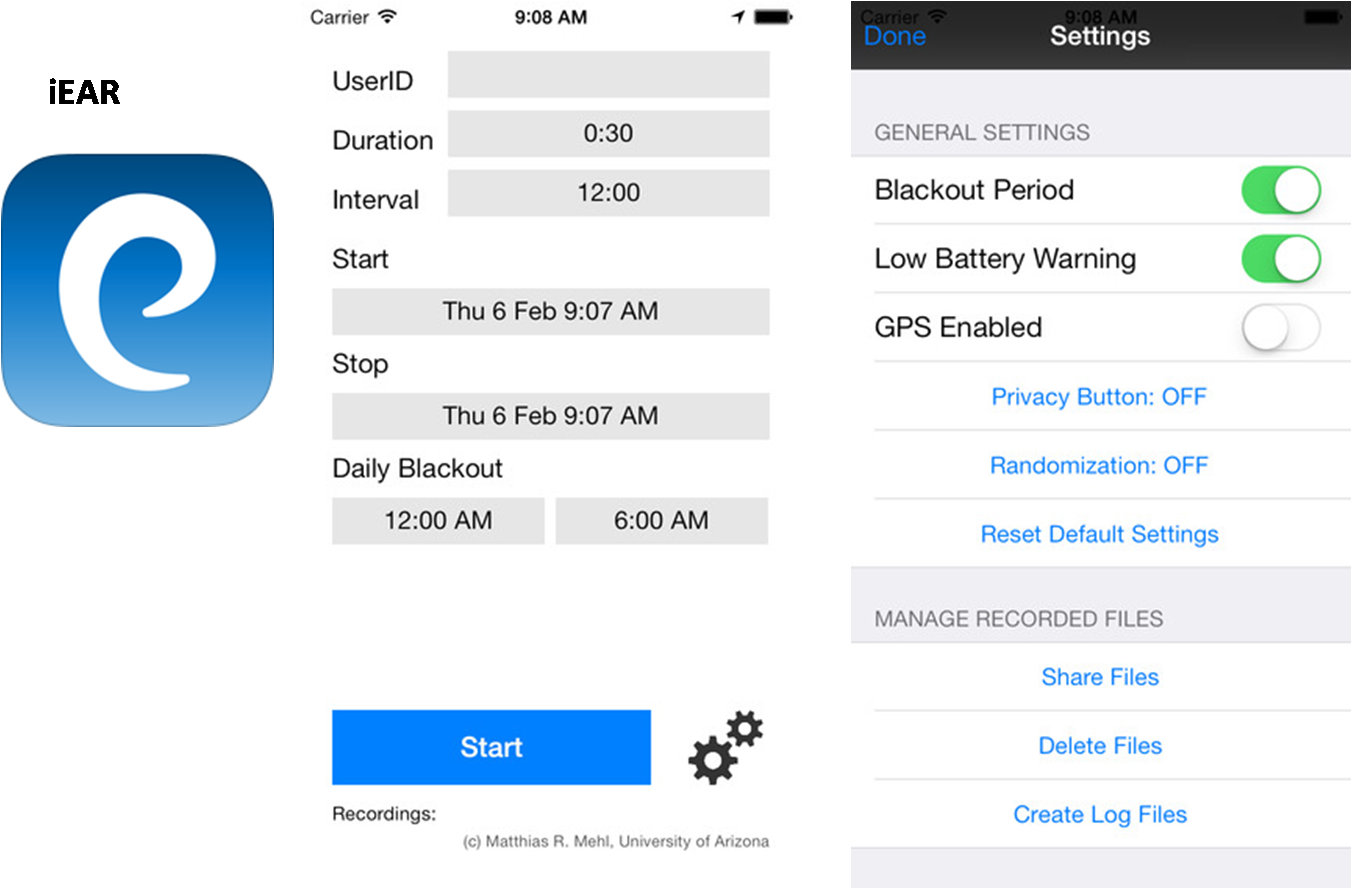}
\caption{The BigEAR app used for recording daily life activities of users in form of daily-audio-log (DAL). It unobtrusively records the ambient sounds  from the surroundings periodically for 50 seconds duration giving out snippets of ambient sounds.}\label{fig_iear}%
\end{figure}
\vspace{-2mm}
\section{Psychological Audio Processing Chain (PAPC)}
Figure~\ref{fig_papc} shows the conceptual view of PAPC framework
proposed in this paper. It is a cascade of audio processing blocks for realizing psychology-based speech processing. The audio data is digitized as a mono-channel signal sampled at 11.025 kHz sampling rate with 16-bit precision. The collected audio data inherently contains noise and significant amount of silences. It necessitated the preprocessing the audio with silence suppression and noise reduction before feeding it to feature extraction and mood classification system. 
\begin{figure}[!t]
\centering
\includegraphics[width=240bp]{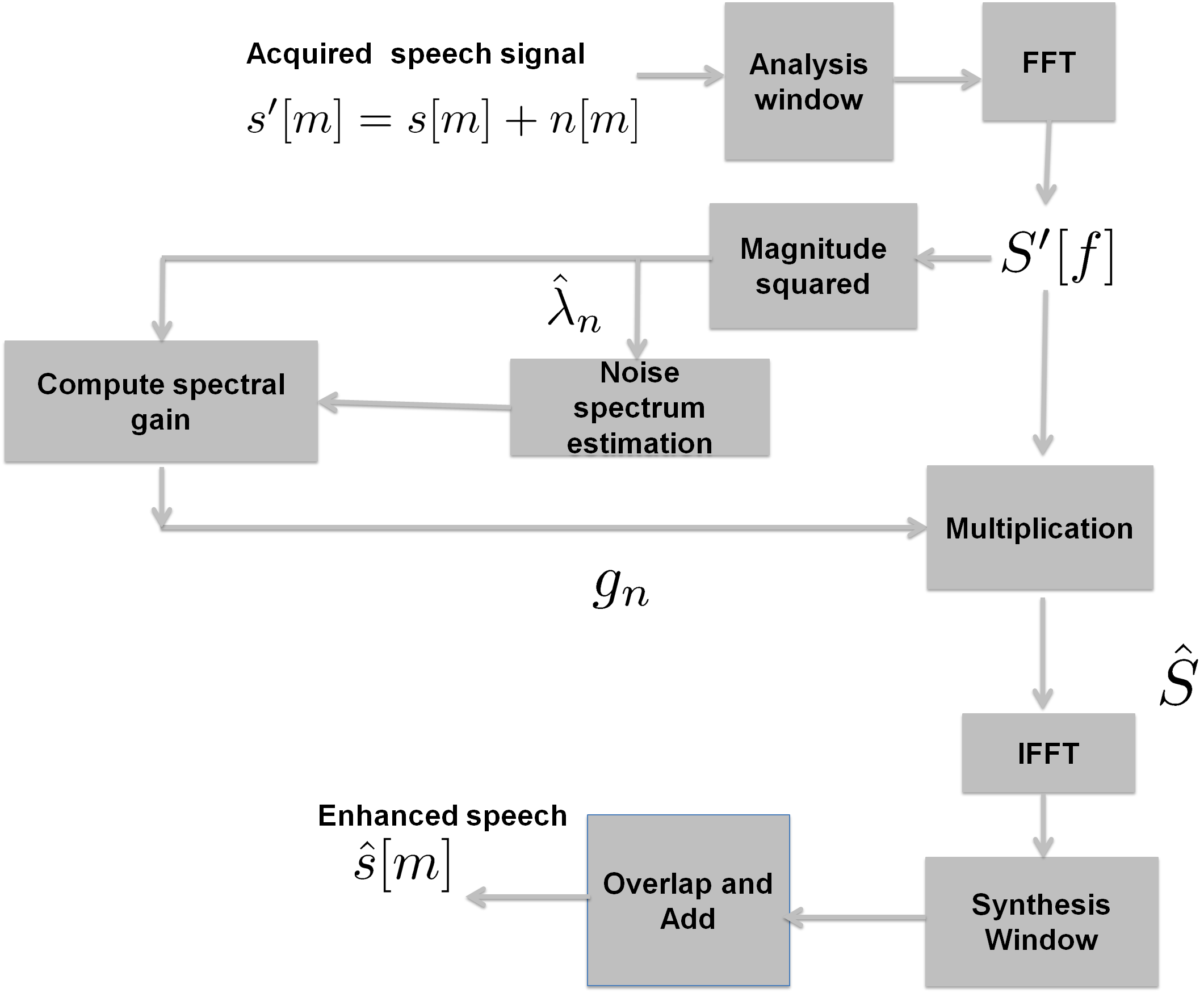}
\caption{Block diagram of the method used for reducing the background noise~\cite{cohen2003noise}.}
\label{fig_noise_algo}%
\end{figure}
For audio processing, we employed overlapping 40ms windows with skip rate of 10ms. 
\vspace{-1mm}
\subsection{Silence Removal}
\begin{figure}[!t]
\centering
\includegraphics[width=240bp]{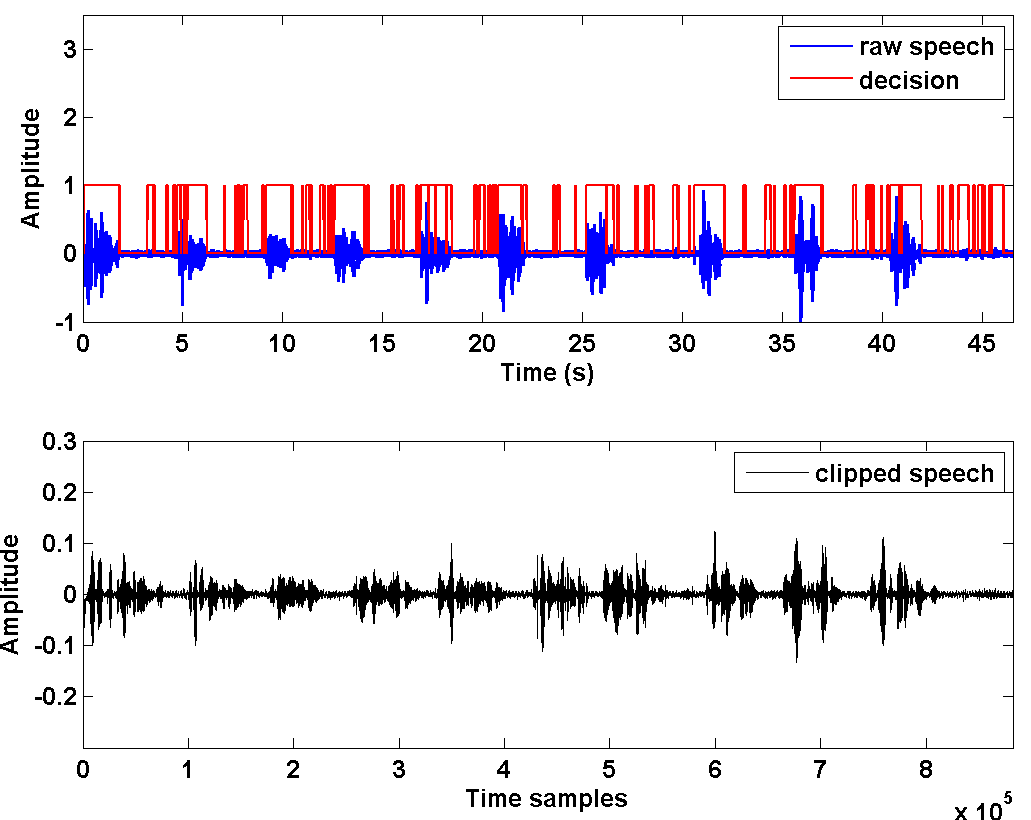}
\caption{The top sub-figure shows the time-domain audio signal and the corresponding speech non-speech decision. The bottom sub-figure shows the separated speech contained in audio recording~\cite{tan2010low}.}
\label{fig_clipped2}%
\end{figure}
The speech non-speech decision was made by the algorithm developed in ~\cite{tan2010low}. It is based on the effective selection of audio-frames. The short time-windows of the speech signal are stationary (for 25-40ms windows). However, for an extended time-duration (more than 40ms), the statistics of the speech signals changes significantly rendering unequal relevance of the speech-frames. It necessitates the selection of effective frames on the basis of posteriori signal to noise ratio (SNR). The authors used the energy distance for measuring the relevance of speech-frames. This method was found to be effective even at low SNRs and had reduced computational complexity as compared to other available algorithms. Hence, we chose it for suppressing the silences in audio recording obtained from iEAR app. Figure~\ref{fig_clipped2} illustrates the speech non-speech decision for an audio signal. 
\vspace{0mm}
\subsection{Noise Reduction}
\begin{figure}[!t]
\centering
\includegraphics[width=240bp]{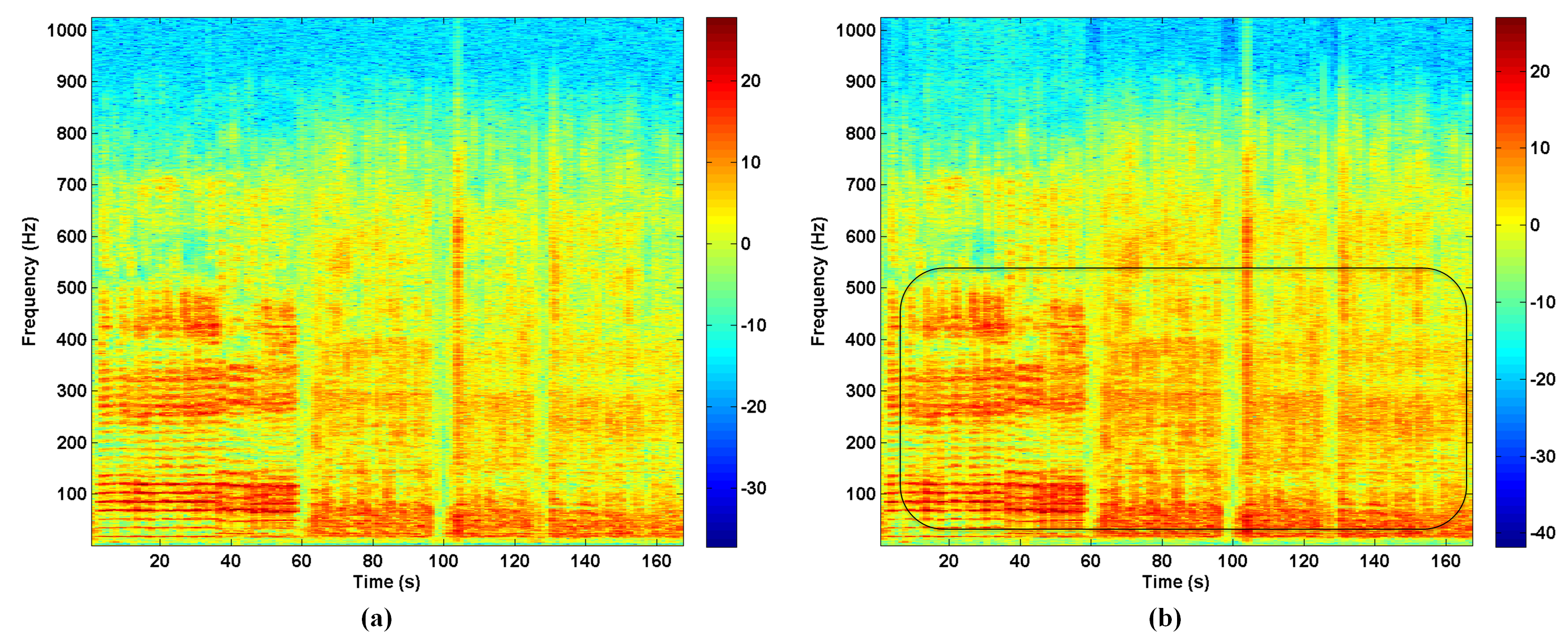}
\caption{The spectrogram of (a) original speech signal and (b) enhanced signal. Increase in energy of certain frequencies marked by squared block is clearly evident.}
\label{fig_spec_both}%
\end{figure}
\vspace{0mm}
The speech signal inevitably contain background noise. Authors in ~\cite{cohen2003noise} developed a method for reducing non-stationary noise in audio signal. It optimized log-spectral amplitude of speech signal using noise estimates from minima controlled recursive averaging. Two functions were used for accounting the probability of speech presence in various sub-bands. One of these functions was based on the time frequency distribution of the a-priori SNR and was used for estimation of the speech spectrum. Second function was decided by the ratio of energy of noisy speech-segment and its minimum value within the time-window. Figure~\ref{fig_noise_algo} shows the block diagram of the noise reduction method proposed in ~\cite{cohen2001speech},~\cite{cohen2003noise}. Figure~\ref{fig_spec_both} shows the spectrogram of speech signal and that of corresponding enhanced signal. Energy enhancement is clearly visible in the frequency range 10-500 Hz that is marked by the square block. 
Let $ S[n] $ and  $ S'[n] $ be the speech before and after noise reduction, we can derived a metric, change in signal-to-noise energy (CSNE) in dB as follows:
\begin{equation}
CSNE_{dB} = \frac{\sum_{n=0}^{L-1}S'[n]^2}{\sum_{n=0}^{L-1} \left(S'[n]-S[n] \right)^2}
\end{equation}
where $ L $ is the length (in samples) of the speech signals and $ n $ is the time-index. CSNE in dB can be used for environment classification as discussed in next sections. Usually, outdoors are accompanied by high CSNE values as compared to indoors. The TV/music gave intermediate values for CSNE.  
\vspace{0mm}
\subsection{Speaker Segmentation in Speech Signal}
The audio signal contains voices of various speakers. Before making inference on moods of each one of the speaker such as couples, we have to separate the speaker segments in audio signal. We used the method developed
in~\cite{giannakopoulos2012fisher} for speakers diarization of enhanced speech signal after noise reduction as shown in Figure~\ref{fig_papc}. Authors used Mel-frequency Cepstral Coefficient (MFCC) features for finding
Fisher near-optimal linear discriminant subspace for developing an unsupervised method for speaker segmentation. The proposed approach relies on a semi-supervised version of Fisher linear discriminant analysis (FLD) and utilizes the sequential structure of speech signal. The overall accuracy of 92 \% with respect to ground-truth was obtained for speaker diarization using this technique on our data. Readers are referred to~\cite{giannakopoulos2012fisher} for more details on implementation strategy.
\subsection{Classification of Acoustic Environment}
The background-noise can quantify the acoustic environment where social conversation take place. For example, on an average outdoors are usually more noisy as compared to indoors. In this paper, we have used change in signal-to-noise energy (CSNE) in dB and speech signal as features in a decision tree classifier for classification of environment into three categories namely, indoor, outdoor, and TV/music. These features were computed on short overlapping frames of 40 ms with 10 ms skip rate. All the frame-level features were cascaded into a feature vector and fed to a C4.5 decision-tree classifier~\cite{quinlan1993c4}. We achieved an overall accuracy of 83.68\% using these features and decision tree classifier.
\vspace{-2mm}
\section{Experiments}
\subsection{Smart Audio Data}
The participants were recruited by the Department of Psychology in University of Arizona for collection of audio data. All the participants signed the consent form approved by the University of Arizona Institutional Review Board before collecting the psychological audio data. The individuals using iEAR were couples coping with breast cancer. The iEar app running on iPhones collected short chunks of audio data (50 seconds) at random instances throughout the day, thus, capturing the acoustic scenarios at irregularly spaced time intervals~\cite{iearapp}. Each day accumulated 80 files on average from each participant. We used the data from 10 individuals who used iEAR app over 1 week, thus a total of 5600 audio recording, each one of 30 seconds duration. Manual transcription was obtained for each of these files along with mood labels. We divided the corpus into two parts: training set and testing set. Out of total, 5600 audio files, 4500 was used for training the classification system and rest 1100 for testing. Same division of data was used for determining the accuracy of speech activity detection. 
%
\begin{table*}[t]
\centering
\caption{Emotion categories and accuracy in detection}
\begin{tabular}{*{3}{|c|}}
\hline
S.No. & Moods category& Physical Interpretation \\
\hline
1& laugh
&  situation where either primary or secondary speaker is laughing  \\
\hline
2&	sing & either of speakers are singing or from TV/surroundings\\
\hline
		3& cry & either of speakers are singing or from TV/surroundings	\\
		\hline
		4& arguing & between couples	\\
		\hline
		5& sigh& from either of speakers or from TV/surroundings\\
		\hline
	\end{tabular}
	\label{table_task}
\end{table*}
\vspace{0mm}
\subsection{Perceptual Audio Features}
We designed some new features for our task and fused it with the state-of-the-art features for emotion detection from speech and music. The Munich Versatile and Fast Open-Source Audio Feature Extractor (OpenSMILE) is a feature extractor for emotion recognition.  We used the default parameter set in v2.1 openSMILE for research discussed in this paper. OpenSMILE had been successfully applied for emotion identification in music, speech and other mixed sources,~\cite{eyben2010music}. Basically, the OpenSMILE features consists of sound, speech, and music features along with cross-domain features. The energy, spectral and voicing related low-level descriptors (LLDs) are used. We processed the audio files for noise reduction and suppression of silences before feeding it to openSMILE feature extractor. The features were observed to be correlated that necessitated preprocessing with DCT before feeding it into classification system. Next, we used DCT for decorrelating the features set (6373 dimensional features from OpenSMILE), thus reducing the dimension to 3000. Later we appended our curated feature set to OpenSMILE post-processed features. 
\vspace{0mm}
\subsubsection{Jitter}
Jitter ($J_{1} $) quantifies cycle-to-cycle changes in the
vocal period. Fundamental frequency contour was used to
obtain the two features namely absolute and relative jitter ~\cite{tsanas2012accurate}. $J_{1} $ is defined as the average absolute difference between consecutive periods. Mathematically, it is given as:
\begin{equation}
J_{1}= \frac{1}{M}\sum_{j=1}^{M-1} \vert F_{j} - F_{j+1} \vert
\end{equation}
where $F_{j} $ is the j-th extracted vocal period and $M$ is the number of extracted vocal periods.
\textit{Relative Jitter}, $ J_{2} $, is defined as the ratio of average absolute difference between consecutive vocal periods to the
average period. Mathematically, it is given by (expressed as percentage):
\begin{equation}
J_{2}= \frac{\frac{100}{M}\sum_{j=1}^{M-1} \vert F_{j} - F_{j+1} \vert}{\frac{1}{M}\sum_{j=1}^{M-1} \vert F_{j} \vert}
\end{equation}
\vspace{-1mm}
\subsubsection{Shimmer}
It measures variations in intensity of the speech signal. It captures the instabilities in oscillating pattern of the vocal folds that influences the cycle-to-cycle changes in speech amplitude. It is defined  as average absolute difference in intensities from two successive cycles. Shimmer in decibels (dB) is mathematically, given as: 
\begin{equation}
S_{dB}= \frac{20}{M} \sum_{j=1}^{M-1}\vert \log10 \frac{A_{i}}{A_{i+1}} \vert
\end{equation}
where $ A_{i} $ is the maximum amplitude of speech in i-th frame and M is the number of frames. 
\vspace{0mm}
\subsubsection{Frequency Modulation}
It quantifies the presence of sub-harmonics in speech signal. There are no references for its use in emotion classification. Usually, speech signals with many sub-harmonics lead to a more complicated interplay between various harmonic components making it relevant for emotion identification. Mathematically, it is given as~\cite{sun2000pitch}:
\begin{equation}
F_{mod}= \frac{ \max \left( F_{j}\right)_{j=1}^{M} - \min \left( F_{j}\right)_{j=1}^{M}}{\max \left( F_{j}\right)_{j=1}^{M} + \min \left( F_{j}\right)_{j=1}^{M}
}
\end{equation}
where $F_{mod}$ is frequency modulation, and $ F_{j} $ is the fundamental frequency of j-th speech frame. 
\vspace{0mm}
\subsubsection{Frequency Range}
The range of frequencies is an important feature of speech signal that quantifies its quality and emotions associated with it~\cite{banse1996acoustic}. We computed the frequency range as the difference between 5-th and 95-th percentiles. Mathematically, it becomes: 
\begin{equation}
F_{range}= F_{95\%} - F_{5\%} 
\end{equation}
Taking 5-th and 95-th percentiles helps in eliminating the influence of outliers in estimates of fundamental frequency. 
\vspace{0mm}
\subsubsection{Harmonics to Noise Ratio}
Harmonics to Noise Ratio (HNR) quantifies the noise present in the speech signal that results from incomplete closure of the vocal folds during speech production process~\cite{tsanas2012accurate}. In this paper, we have used method proposed in~\cite{boersma1993accurate} for HNR estimation. The average and standard deviation of the segmental HNR values are later used for emotion prediction. Lets assume that $ R_{xx} $  is normalized autocorrelation and $ l_{\max} $ is the lag (in samples) at which it is maximum, except the zero lag. Then, HNR is mathematically given by~\cite{boersma1993accurate}:   
\begin{equation}
HNR_{dB}=  10 \log10 \left( \frac{ R_{xx}(l_{\max}}{ 1- R_{xx}(l_{\max}} \right)
\end{equation}
%
\vspace{-1mm}
\subsubsection{Spectral Centroid} It is the \textit{center of mass}  of spectrum. It measure the \textit{brightness} of an audio signal.  Spectral centroid of a spectrum-segment is given by average values of frequency weighted by amplitudes, divided by the sum of amplitudes. Mathematically, we have  
\begin{equation}
SC=\frac{\sum _{n=1}^{N} k F[k]}{\sum _{n=1}^{N} F[k]}
\end{equation}
where SC is the spectral centroid, and $ F[k] $ is amplitude of k-th frequency bin of discrete Fourier transform of speech signal. Average spectral centroid for complete audio signal was used for emotion identification~\cite{paliwal1998spectral}. 
\vspace{0mm}
\subsubsection{Spectral Flux}
It quantifies the rate of change in power spectrum of speech signal. It is calculated by comparing the normalized power spectrum of a speech-frame with that of other frames. It determines the timbre of speech signal. 
Spectral flux had been previously used for emotion recognition in music~\cite{yang2008regression}.
\vspace{0mm}
\subsubsection{Spectral Entropy}
Spectral Entropy was used for emotion recognition in speech~\cite{lee2008speech}. It is given by:
\begin{equation}
SE=\frac{-\sum P_{j} log(P_{j})}{log(M)}
\end{equation}
where SE is the spectral entropy, $ P_{j} $ is the power of j-th frequency-bin and M is the number of frequency-bins. Here, $\sum P_{k} = 1 $ as the spectrum is normalized before computing the spectral entropy. 
\vspace{0mm}
\subsubsection{Spectral Flatness} 
It measures the flatness of speech power spectrum. It quantifies how similar the spectrum is to that of a noise-like signal or a tonal signal. Spectral Flatness (SF) of white noise is 1 as it has constant power spectral density. A pure sinusoidal tone has SF close to zero showing the high concentration of power at a fixed frequency. Mathematically, SF is ratio of geometric mean of power spectrum to its average value~\cite{johnston1988transform}.
\vspace{0mm}
\subsubsection{Sharpness}
Sharpness is a mathematical function that quantifies the sensory pleasantness of a speech signal. High sharpness implies low pleasantness. It value depends on the spectral envelope of signal, amplitude level and its bandwidth. The unit of sharpness is acum (Latin expression). The reference sound producing 1 acum is a narrowband noise, one critical band wide with 1 kHz center frequency at 60 dB intensity level~\cite{fastl2007psychoacoustics}. Sharpness, $S$ is mathematically defined as
\begin{equation}
S= 0.11 \frac{\sum_{0}^{24 Bark}L_{0} \cdot g(z) \cdot z \cdot dz}{\sum_{0}^{24 Bark}L_{0}\cdot dz} acum
\label{eqn_sharpness}
\end{equation}
However, its  numerator is weighted average of specific loudness ($L_{0}$)
over the critical band rates. The weighting function, $g(z)$, depends on critical band rates. The $g(z)$ could be interpreted as the mathematical model for the sensation of sharpness. 
%
\vspace{-2mm}
\section{Results \& Discussions}
The 11 dimensional curated features consisting of absolute and relative jitter, shimmer, frequency modulation, frequency range, Harmonic-to-noise ratio,spectral centroid, spectral flux,  spectral entropy, spectral flatness and sharpness are cascaded with 3000 dimensional DCT-processed OpenSMILE features. These final features of dimension 3011 are later fed to a deep neural network(DNN) with 4 hidden layer with 2048, 2048, 1024 and 1024 nodes respectively. The input node had 3011 dimension while output node had 5 nodes. We used log-sigmoid non-linearity in hidden layers and soft-max linearity of output layer. The cross-entropy between ground-truth and output was used as objective function in training the DNN. Deep Neural Networks (DNNs) are large scale version of feed-forward neural networks that have been successfully validated for learning the complex functional relations. Use of DNN for detecting emotions from speech and music have been well accounted in literature~\cite{sanchez2014deep}. Each feature dimension was normalized to have zero mean and unit variance before feeding it to DNN. Dropout strategy was used to train the DNN as it avoids over-fitting that can happen in high dimensional neural network with hybrid features from music and speech~\cite{srivastava2014dropout}. We used kaldi toolkit for realizing DNN training and testing as it is popular for training deep neural networks especially in speech recognition community~\cite{povey2011kaldi}. The overall average accuracy of DNN classifier with complete feature set comes out to be 88.76\% that is a competitive accuracy owing to naturalistic noisy conditions where the data recording takes place. It was observed from the collected audio data that the cancer related discussion were more common with couples as compared to friends and family. The discussion were more informational than emotional. In addition, the involvement of couples in discussions lead to be adjustment in cancer patients. The main benefit of using BigEAR is that it provides objective criterion for such studies and evaluation. 
\vspace{-2mm}
\section{Conclusions}
This paper developed the BigEAR framework was validated to be a unobtrusive smartphone-based system that could identify the emotion and ambient correlates during social conversations in naturalistic scenarios. The case study involved couples coping with breast cancer. The BigEAR framework is powered by psychological audio processing chain (PAPC) for inferring the environment where the conversations take place and may contain mood or emotional information of the speakers. The advantage of BigEAR framework is that the psychologists no longer analyze the increasing amount of acoustic big data that demand listening to each and every audio file and classify the moods into categories. The downside is that the BigEAR is just 88.76\% accurate as compared to ground-truth labels from human listeners. However, this accuracy level is in an acceptable range for clinicians and psychologists since BigEAR offers automated audio analysis as opposed to human ratings that are time consuming, costly, and complicated w.r.t. manpower management.  
\vspace{-2mm}
\bibliographystyle{IEEEtran}
\bibliography{4.BigDataWorkshop}
\end{document}